\documentstyle[preprint,revtex]{aps}
\begin{document}
\draft
\begin{title}
Gravitational Laser Back-Scattering
\end{title}
\author{ S.\ F.\ Novaes$^{(1)}$  and  D.\ Spehler$^{(1,2)}$}
\begin{instit} $^{(1)}$ Instituto de F\'\i sica Te\'orica,
Universidade Estadual Paulista\\
Rua Pamplona 145, 01405-900 S\~ao Paulo, Brazil \\
$^{(2)}$ Universit\'e Louis Pasteur, I.U.T. \\
3 rue S\b{t} Paul, 67300 Strasbourg, France
\end{instit}
\begin{abstract}
A possible way of producing gravitons in the laboratory is
investigated. We evaluate the cross section electron + photon
$\rightarrow$ electron + graviton in the framework of linearized
gravitation, and analyse this  reaction considering the photon coming
either from a laser beam or from a Compton  back-scattering process.
\end{abstract}

\pacs{04.30.+x, 04.80.+z}


\newpage

The attempts to detect gravitational radiation usually rely on the
observation of astrophysical objects. These sources, like supernovae
explosions or coalescing binaries, have in general very uncertain
frequency, strength, and spectrum of gravitational waves.
A considerable improvement in the search of the  gravitational waves
would occur if we succeed in producing them in the laboratory.
In this case, the energy and propagation direction could, at least in
principle, be controlled.

In this paper, a mechanism of graviton production in a linear
electron-positron collider is analysed. Inspired by the Compton laser
back-scattering process  \cite{las0}, we suggest that the collision
of laser photons of few electron-volts, at small angle, with an
energetic electron beam is able to generate a very collimated graviton
beam.

We evaluate, in a first step, the cross section for the process
electron + photon $\rightarrow$ electron + graviton, and estimate
the total radiated power assuming that the photons come from a laser
beam.  The scenario where the photons are harder,  as a result of a
Compton back-scattering process, is examined in a second step.

The coupling of the matter fields with gravity can be obtained using
the weak field approximation \cite{pre}. The metric is assumed to be
close to the Minkowski one ($\eta_{\mu\nu}$)
\begin{equation}
g_{\mu\nu} = \eta_{\mu\nu} + \kappa h_{\mu\nu} \; ,
\label{g}
\end{equation}
where $h_{\mu\nu}$ is the graviton field and $\kappa \equiv
\sqrt{32\pi G} = 8.211 \times 10^{-19}$ GeV$^{-1}$ in natural
unities. The vierbein is expanded as
\begin{equation}
{\cal V}^a_{\;\; \mu} = \delta^a_{\;\; \mu} + \frac{\kappa}{2}
h^a_{\;\; \mu} - \frac{\kappa^2}{8} h^a_{\;\; \alpha}
\delta^\alpha_{\;\; b} h^b_{\;\; \mu} + \cdots \; .
\label{tet}
\end{equation}

The above expressions allow us to write the relevant part of the
action as
\begin{eqnarray}
S &=& \int \mbox{d}^4x \: \biggl\{
\frac{\kappa}{2} \biggl[ \left[ -\frac{i}{4} \bar{\psi}  \left(
\gamma^a \partial_\mu + \gamma_\mu \partial^a \right) \psi +
\frac{i}{4} \left(\partial_\mu \bar{\psi} \gamma^a + \partial^a
\bar{\psi} \gamma_\mu \right) \psi \right] h_a^{\;\; \mu}
\nonumber \\
&&+ \left( \frac{i}{2} \bar{\psi} \gamma^\mu \partial_\mu \psi
- \frac{i}{2} \partial_\mu \bar{\psi} \gamma^\mu \psi - m \bar{\psi}
\psi \right) h  \biggr]
- \frac{\kappa}{8} \left(\eta^{\mu\alpha} \eta^{\nu\beta} h -
4 h^{\mu\alpha} \eta^{\nu\beta}\right)
F_{\mu\nu} F_{\alpha\beta}
\nonumber \\
&& - \frac{e \kappa}{4} \left( \delta_a^{\;\; \mu}
h - h_a^{\;\; \mu} \right) \bar{\psi} \left( A_\mu \gamma^a + A^a
\gamma_\mu \right) \psi \biggr\} \; ,
\label{S}
\end{eqnarray}
where we omitted the free electron and photon Lagrangians, and the
usual QED fermion-photon coupling, and neglected the  ${\cal
O}(\kappa^2)$ terms.

In order to evaluate the cross section for the process electron +
photon $\rightarrow$ electron + graviton, we used the Weyl-van der
Waerden spinor  technique to describe the graviton helicity wave
function \cite{our} and the graviton couplings with bosons and
fermions \cite{pre}. The cross section of the reaction  $e^- (p) +
\gamma (q) \rightarrow e^- (p^\prime) + g (k)$ is given by
\begin{eqnarray}
\frac{d\sigma}{dt} &=& \frac{e^2 \kappa^2}{16 \pi} \frac{1}{(s -
m^2)^2}  \frac{(us - m^4)}{t} \biggl[ \left( \frac{m^2}{s - m^2} +
\frac{m^2}{u - m^2} \right)^2 \nonumber \\
&& +  \frac{m^2}{s - m^2} + \frac{m^2}{u - m^2}
- \frac{1}{4} \left( \frac{s - m^2}{u - m^2} + \frac{u - m^2}{s -
m^2} \right) \biggr]
\label{dsig:dt}
\end{eqnarray}

This expression can be written in a more compact form if we
define new variables,
\begin{eqnarray}
x &=& \frac{s - m^2}{m^2} = \frac{2 E q_0}{m^2} (1 + \beta \cos
\alpha) \; ,  \nonumber \\
y &=& \frac{- t}{s - m^2} =
\frac{k_0}{E} \frac{1 + \cos(\theta - \alpha)}{1 + \beta  \cos
\alpha }  \; ,
\label{xy}
\end{eqnarray}
where $m$, $E$, and $\beta = (1 - m^2/E^2)^{1/2}$ are respectively
the electron mass, energy, and velocity in the laboratory frame. The
laser, of energy $q_0$,  is supposed to make an angle $\alpha$ with the
electron beam. The scattered graviton has energy $k_0$,
\begin{equation}
k_0 = q_0 \frac{1 + \beta \cos \alpha}{1 - \beta \cos \theta +
(q_0/E)  \left[ 1 + \cos (\theta - \alpha) \right]}
\label{k0}
\end{equation}
where $\theta$ is the angle between the graviton and the incoming
electron. Note that, for $\alpha \simeq 0$, the variable $y$
represents the fraction of the electron energy carried by the
graviton in the forward direction ($\theta = 0$).

Equation (\ref{dsig:dt}) expressed in terms of $x$ and $y$ becomes
\begin{equation}
\frac{d\sigma}{dy} = \frac{e^2 \kappa^2}{64 \pi}
\left( \frac{1}{y} - \frac{x + 1}{x} \right) F(x,y) \; ,
\label{dsig:dy}
\end{equation}
with
\begin{equation}
F(x,y) \equiv \left[ 1 - y + \frac{1}{1 - y} - \frac{4 y}{x (1 - y)}
+  \frac{4 y^2}{x^2 (1 - y)^2} \right] \; .
\label{F}
\end{equation}
Equation (\ref{dsig:dy}) shows that the cross section for large values
of $y$ is suppressed by an overall factor ($1/y - 1/y_{max}$)  which
makes the  distribution very sharp for small values of $y$. This is a
major difference from the  Compton cross section $e + \gamma_\ell
\rightarrow e + \gamma_b$,  where $\gamma_{\ell (b)}$ is the laser
(back-scattered) photon. In the latter case, the $y$-distribution
assumes its maximum value for $y = y_{max} = x/(x+1)$, giving rise to a
hard photon spectrum.

Let us start by analysing our result when the initial photon
originates from a laser beam.  We consider four different designs of
$e^+e^-$ linear  colliders \cite{pal}:  SLC ($E = 50$ GeV, ${\cal L}
=  5 \times 10^{29}$ cm$^{-2}$ s$^{-1}$),   Palmer-G ($E = 250$ GeV,
${\cal L} = 5.85 \times 10^{33}$ cm$^{-2}$ s$^{-1}$),
Palmer-K ($E = 500$ GeV, ${\cal L} = 11.1 \times 10^{33}$
cm$^{-2}$ s$^{-1}$), and VLEPP ($E = 1000$ GeV, ${\cal L} = 10^{33}$
cm$^{-2}$ s$^{-1}$).

Assuming $q_0 = 1$ eV, and $\alpha=0$, we compare the behavior of the
graviton  angular distribution  $d\sigma/d\cos\theta$ for different
electron beam energies. The angular distribution is peaked for very
small angles, close to the electron direction [ Fig. (\ref{fig:1})].
By changing the value of $\alpha$ we are of course able to shift this
peak away from the electron beam.

Let us suppose that the laser has flash energy of $2.5$ J, the same
repetition rate as the electron pulse frequency, and a rms radius of
$20 \mu$m. In this case, the electron laser luminosity is  ${\cal
L}_{e\gamma_\ell} = \eta {\cal L}$, ${\cal L}$ being the
collider luminosity, and  $\eta = A_e N_{\gamma_\ell} /
A_{\gamma_\ell} N_e$. Here, $A_{e(\gamma_\ell)}$ is the electron
(laser) beam cross  section and  $N_{e(\gamma_\ell)}$ is the number
of electrons (photons) in the bunch (flash).

The radiation power emitted in solid angle of semi-angle ($\theta_0$)
due to the graviton emission is
\[
P_\ell(\theta_0) = {\cal L}_{e\gamma_\ell} \;
\int_{0}^{\theta_0} d\theta \; \sin\theta
\frac{d\sigma}{d\cos\theta}(\theta) \; k_0(\theta)
\]

In Fig. \ref{fig:2} we show the spectral distribution of the emitted
radiation
\[
\frac{dP_\ell}{dk_0} \propto \frac{k_0}{\beta E - q_0}
\left(\frac{\beta E - q_0}{k_0 - q_0} - \frac{x + 1}{x}  \right)
F\left(x,\frac{k_0 - q_0}{\beta E - q_0} \right)  \; .
\]
We note that this distribution is maximum for $k_0 = k_0^{min} =
q_0$, and is almost independent of the electron beam energy.

Another way of obtaining the reaction $e + \gamma  \rightarrow e +
g$ is taking advantage of the energetic photons produced by the
Compton back-scattered process. The energy spectrum of back-scattered
photons is \cite{las0,las}
\begin{equation}
{\cal F}_{\gamma/e} (x,z) \equiv \frac{1}{\sigma_c}
\frac{d\sigma_c}{dz} =  \frac{1}{C(x)} F(x,z)
\label{F:laser}
\end{equation}
where $\sigma_c$ is the Compton cross section, and  $z = q_\gamma/E$
is the fraction of the  initial electron carried by the photon. The
function $F(x,z)$ was defined in Eq.(\ref{F}), with $x$ given by
Eq.(\ref{xy}), and
\begin{equation}
C(x) = \left(1 - \frac{4}{x} - \frac{8}{x^2}  \right) \ln (1 + x) +
\frac{1}{2} + \frac{8}{x} - \frac{1}{2(1 + x)^2}
\label{C}
\end{equation}

Defining the ratio of electron-photon invariant mass squared
($\hat{s} = 4Eq_\gamma$) to invariant mass squared of the collider,
\[
\tau \equiv \frac{\hat{s}}{s} = \frac{q_\gamma}{E} \; ,
\]
and assuming a conversion coefficient (the average number of
converted photons per electron) $\xi = 0.65$ \cite{tel}, the
$e\gamma_b$ luminosity can  be written in terms of the machine
luminosity as ${\cal L}_{e\gamma_b} = \xi {\cal L}$. The luminosity
distribution of the back-scattered photons is
\begin{equation}
\frac{d {\cal L}}{d\tau} = \xi  \; {\cal L} \;  {\cal F}_{e/\gamma}
(x,\tau)
\label{l:eg}
\end{equation}

In this case, the radiation power, emitted in solid angle of
semi-angle ($\theta_0$), is
\[
P_b(\theta_0) =
\int_{0}^{\tau_{max}} d\tau \frac{d{\cal L}}{d\tau}
\int_{0}^{\theta_0} d\theta \; \sin\theta
\frac{d\hat{\sigma}}{d\cos\theta}(\theta,\tau) \; k_0(\theta,\tau)
\]
where $\tau_{max} = x/(x+1)$, and $\hat{\sigma}$ is the elementary
cross section for the process $e + \gamma  \rightarrow e + g$
evaluated at $s = \hat{s}$.

In Table \ref{tab1} we present the total radiation power for
different values of $\theta_0$, and for a laser energy  ($q_0$) of
$1$ eV.  Since $x$ increases when we increase either the laser energy
or the electron beam energy, the different choices of colliders
considered here give a reasonable idea of the laser back-scattering
capabilities to produce gravitons in the laboratory.

As pointed out before, gravitons can be produced  via $e +
\gamma_\ell \rightarrow e + g$, where  $\gamma_\ell$ is a laser
photon, and via $e + e (+ \gamma_\ell) \rightarrow e + \gamma_b
\rightarrow e + g$, where $\gamma_b$ is the laser back-scattered
photon. In the former case, the radiation power $P_\ell$ is almost
independent of the particular choice of $\theta_0$, since the angular
distribution of the graviton is peak at very small angles
relative to the electron beam direction. On the other hand, in the
latter case, the back-scattered photon is very hard, giving
rise to a flatter angular distribution and making the  dependence of
$P_b$ on $\theta_0$ more evident. The total forward emitted power
becomes, therefore, small for hard photons.

In a recent paper, Chen \cite{che} has examined the production of
gravitons due to the interaction between the beams near the
interaction point in an $e^+e^-$ collider (gravitational
beamstrahlung). His results for the total radiation power, at SLC,
is roughly one order of magnitude bellow the one for the gravitational
laser back-scattering process. For the next generation of linear
$e^+e^-$ colliders, the gravitational beamstrahlung yields a larger
radiation power. Nevertheless, the coherent contribution is again
smaller than the one coming from the laser back-scattering.

We should stress that, in spite of the small emitted radiation power,
the gravitational laser back-scattering process yields a very
collimated beam of gravitational waves with a defined
energy spectrum (see Fig. \ref{fig:2}). These points could be  of
great help for the complex issue of gravitational wave detection
\cite{bla}.

\acknowledgments
We are very grateful to R.\ Aldrovandi, O.\ J.\ P.\ \'Eboli,
and  C.\ O.\ Escobar for useful discussions. This work was partially
supported by Conselho Nacional de  Desenvolvimento Cient\'\i fico e
Tecnol\'ogico, CNPq (Brazil).

\figure{The angular distribution $d\sigma/\cos\theta$ in arbitrary
units versus the angle in radians for the process $e^- + \gamma_\ell
\rightarrow e^- + g$, where $\gamma_\ell$ comes from a laser beam.
We assumed $q_0$ = 1 eV and plot for different electron beam
energies: E = 50 GeV (dashed), 250 GeV (solid), 500 GeV (dotted),
and 1000 GeV (dot-dashed).
\label{fig:1}}

\figure{Spectral distribution of the graviton radiation power
$dP_\ell/dk_0$ in arbitrary units as a function of $k_0$, for $E =
50$ GeV. \label{fig:2}}

\newpage

\begin{table}
\caption{Total graviton radiation power emitted in a solid angle of
semi-angle $\theta_0$. $P_\ell$ refers to the direct laser photon
process and $P_b$ to the back-scattered photon one. We consider in
both cases a laser photon with energy of 1 eV}
\begin{tabular}{c|c|lc}
Collider & $P_\ell$ (eV/s)     &$\theta_0$  & $P_b$ (eV/s) \\
\tableline
      &                        & $10^0$ &  $1.69 \times 10^{-28}$ \\
SLC   & $1.17 \times 10^{-20}$ & $20^0$ &  $4.48 \times 10^{-28}$ \\
      &                        & $30^0$ &  $7.50 \times 10^{-28}$ \\
\tableline
      &                        & $10^0$ &  $5.02 \times 10^{-24}$ \\
Palmer-G & $2.32 \times 10^{-19}$ & $20^0$ &  $1.47 \times 10^{-23}$\\
      &                        & $30^0$ &  $2.66 \times 10^{-23}$ \\
\tableline
      &                        & $10^0$ &  $1.51 \times 10^{-23}$ \\
Palmer-K & $6.70 \times 10^{-19}$ & $20^0$ &  $4.57 \times 10^{-23}$\\
      &                        & $30^0$ &  $8.47 \times 10^{-23}$ \\
\tableline
      &                        & $10^0$ &  $2.26 \times 10^{-24}$ \\
VLEPP & $1.53 \times 10^{-18}$ & $20^0$ &  $6.94 \times 10^{-24}$ \\
      &                        & $30^0$ &  $1.32 \times 10^{-23}$ \\
\end{tabular}
\label{tab1}
\end{table}

\end{document}